\documentclass[10pt,a4paper,headsepline]{scrbook}

\usepackage{html}
\usepackage{makeidx}
\usepackage{xbooktk}
\usepackage{amsbsy}            
\usepackage{amssymb}
\usepackage{amsxtra}
\usepackage{amsmath}
\usepackage{longtable}
\usepackage{scrpage}
\usepackage{harvard}
\usepackage{hyperref}
\usepackage{graphicx}
\usepackage{color}

\newdimen\defpicwidth
\newdimen\defepswidth
\def\ineps#1#2{\includegraphics[width=#1\defepswidth]{#2.ps}}

\def\SKparam#1#2{}

\long\def\@makecaption#1#2{%
  \vskip\abovecaptionskip
  \sbox\@tempboxa{#1. #2}%
  \ifdim \wd\@tempboxa >\hsize
    {\centerline{\renewcommand{\baselinestretch}{0.8}%
\small\normalsize\parbox{0.8\textwidth}{#1. #2}}}\par
  \else
    \global \@minipagefalse
    \hbox to\hsize{\hfil\box\@tempboxa\hfil}%
  \fi
  \vskip\belowcaptionskip}

{\begin{longtable}{|p{1mm}p{3mm}p{0.8\textwidth}p{1mm}|}
\hline\hline
&&&\\[-4mm]
&& \hskip-20pt \centerline{\bf \summaryname}&\\[1mm]
\hline\hline\endfirsthead
\hline\hline
&&&\\[-4mm]
&& \hskip-20pt \centerline{\bf \summarycontname }&\\[1mm]
\hline\hline\endhead
\hline\hline\endfoot
}
{\end{longtable}}

\gdef\xauthor#1{%
\vspace*{-8mm}%
\begin{flushright}%
  {\sl #1}
\end{flushright}%
\vspace*{3mm}\addtocontents{toc}{\protect\contentsline{}{\sl \protect{#1}}{}{}}
} 

\newcommand{\clink}[2][xpl]{%
\nopagebreak[4]\vspace*{-4mm}\hfill{\ineps{0.05}{#1fig}\,\href{\dircodes{\bookname}/#2.html}{{\tt #2}}}\vspace*{2mm}\newline%
}

\providecommand{\pdf}[1]{}
\def\infig#1#2{\includegraphics[width=#1\defpicwidth]{#2.ps}}

\newcommand{\dirxplore}[1]{../#1}
\newcommand{\dircodes}[1]{\wwwebook/codes/#1}

\def\wwwebook{http://www.quantlet.de}

\def\corollaryname{COROLLARY}
\def\definitionname{DEFINITION}
\def\examplename{EXAMPLE}
\def\exercisename{EXERCISE}
\def\lemmaname{LEMMA}
\def\proofname{PROOF}
\def\propositionname{PROPOSITION}
\def\remarkname{REMARK}
\def\summarycontname{Summary (continued)}
\def\summaryname{Summary}
\def\theoremname{THEOREM}

\def\bookname{}

\gdef\ingif#1#2{}

\areaset{150mm}{220mm}
\defpagestyle{xplorebook}%
{(0pt,0pt){\pagemark\hfill\headmark}{\headmark\hfill\pagemark}{}(\textwidth,0.5pt)}%
{(0pt,0pt){}{}{}(0pt,0pt)}
\defpagestyle{plain}%
{(0pt,0pt){}{}{}(0pt,0pt)}%
{(0pt,0pt){}{}{}(0pt,0pt)}
\DeclareMathOperator{\Bessel}{BES}              
\DeclareMathOperator{\divergence}{div}
\DeclareMathOperator{\grad}{grad}
\DeclareMathOperator{\e}{e}

\newcommand{\dd}{\:\text{d}}
\newcommand{\E}{\mathbb{E}}	

\citationstyle{dcu}
\makeindex

\newdimen\defpicwidth
\begin{document}

\setcounter{page}{1}


\setcounter{chapter}{1}

\chapter*{FX smile in the Heston model\protect\footnote{Chapter prepared for the 2nd edition of Statistical Tools for Finance and Insurance, P.Cizek, W.H{\"a}rdle, R.Weron (eds.), Springer-Verlag, \emph{forthcoming in 2011}.}}

\xauthor{Agnieszka Janek\protect\footnote{Institute of Mathematics and Computer Science, Wroc{\l}aw University of Technology, Poland}, 
Tino Kluge\protect\footnote{MathFinance AG, Waldems, Germany}, 
Rafa{\l} Weron\protect\footnote{Institute of Organization and Management, Wroc{\l}aw University of Technology, Poland}, 
and Uwe Wystup\protect\footnote{MathFinance AG, Waldems, Germany}}


{\small
\textbf{Abstract:}
The Heston model stands out from the class of stochastic volatility (SV) models mainly for two reasons. Firstly, the process for the volatility is non-negative and mean-reverting, which is what we observe in the markets. Secondly, there exists a fast and easily implemented semi-analytical solution for European options. In this article we adapt the original work of Heston (1993) to a foreign exchange (FX) setting. We discuss the computational aspects of using the semi-analytical formulas, performing Monte Carlo simulations, checking the Feller condition, and option pricing with FFT. In an empirical study we show that the smile of vanilla options can be reproduced by suitably calibrating three out of five model parameters.

\textbf{Keywords:}
Heston model; vanilla option; stochastic volatility; Monte Carlo simulation; Feller condition; option pricing with FFT
}

\xsection{Introduction}{hes_intro}{}

The universal benchmark for option pricing is flawed. The Black-Scholes formula is based on the assumption of a geometric Brownian motion (GBM) dynamics with constant volatility. Yet, the model-implied volatilities for different strikes and maturities of options are not constant and tend to be smile shaped (or in some markets skewed).\index{volatility smile}\index{volatility skew} Over the last three decades researchers have tried to find extensions of the model in order to explain this empirical fact. 

As suggested already by \citeasnoun{mer:73}, the volatility can be made a deterministic function of time. While this approach explains the different implied volatility levels for different times of maturity, it still does not explain the smile shape for different strikes. \citeasnoun{dup:94}, \citeasnoun{der:kan:94}, and \citeasnoun{rub:94} came up with the idea of allowing not only time, but also state dependence of the volatility coefficient, for a concise review see e.g. \citeasnoun{fen:05}. This local (deterministic) volatility approach yields a complete market model. It lets the local volatility surface to be fitted, but it cannot explain the persistent smile shape which does not vanish as time passes. Moreover, exotics cannot be satisfactorily priced in this model. 

The next step beyond the local volatility approach was to allow the volatility to be driven by a stochastic process. 
The pioneering work of \citeasnoun{hul:whi:87}, \citeasnoun{ste:ste:91}, and \citeasnoun{hes:93} led to the development of stochastic volatility (SV) models,\index{stochastic volatility (SV)} for reviews see \citeasnoun{fou:pap:sir:00} and \citeasnoun{gat:06}. These are multi-factor models with one of the factors being responsible for the dynamics of the volatility coefficient. Different driving mechanisms for the volatility process have been proposed, including GBM and mean-reverting Ornstein-Uhlenbeck type processes. 

The Heston model stands out from this class mainly for two reasons. Firstly, the process for the volatility is non-negative and mean-reverting, which is what we observe in the markets. Secondly, there exists a fast and easily implemented semi-analytical solution for European options. This computational efficiency becomes critical when calibrating the model to market prices and is the greatest advantage of the model over other (potentially more realistic) SV models. Its popularity also stems from the fact that it was one of the first models able to explain the smile and simultaneously allow a front-office implementation and a valuation of many exotics with values closer to the market than the Black-Scholes model. 
Finally, given that all SV models generate roughly the same shape of implied volatility surface and have roughly the same implications for the valuation of exotic derivatives \cite{gat:06}, focusing on the Heston model is not a limitation, rather a good staring point. 

This chapter is structured as follows. In Section \ref{hes_model} we define the model and discuss its properties, including marginal distributions and tail behavior. Next, in Section \ref{hes_option} we adapt the original work of \citeasnoun{hes:93} to a foreign exchange (FX) setting. We do this because the model is particularly useful in explaining the volatility smile found in FX markets; in equity markets the typical volatility structure is a strongly asymmetric skew (also called a smirk or grimace). In Section \ref{hes_Calibration} we show that the smile of vanilla options can be reproduced by suitably calibrating the model parameters. Finally, in Section \ref{hes_beyond} we briefly discuss the alternatives to the Heston model.

\xsection{The model}{hes_model}{}

\begin{figure}[tbp]
\begin{center}
\infig{1.4}{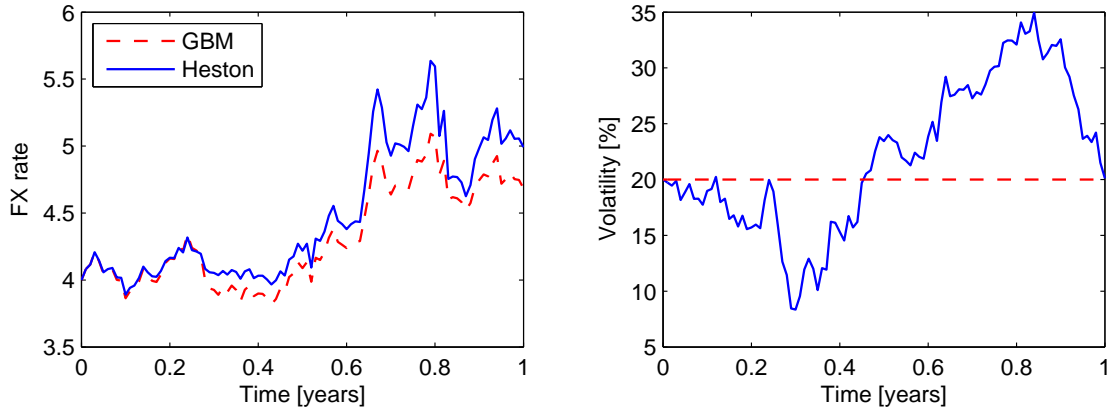}
\caption{Sample trajectories (\textit{left panel}) and volatilities (\textit{right panel}) of the GBM and the Heston spot price process (\protect\ref{eq:heston1}) obtained with the same set of random numbers.}
\label{fig-STF2hes01}
\end{center}
\clink[m]{STFhes01}
\end{figure}

Following \citeasnoun{hes:93} consider a stochastic volatility model with GBM-like dynamics for the spot price:\index{Heston model}
\begin{equation}\label{eq:heston1}
dS_t = S_t\left(\mu\,dt+\sqrt{v_t}dW_t^{(1)}\right)
\end{equation}
and a non-constant instantaneous variance $v_t$ driven by a mean-reverting square root (or CIR) process:\index{process!square root}\index{process!CIR}
\begin{equation}\label{eq:volprocess}
dv_t=\kappa(\theta-v_t)\,dt+\sigma\sqrt{v_t}dW_t^{(2)}.
\end{equation}
The stochastic increments of the two processes are correlated with parameter $\rho$, i.e. $dW_t^{(1)}dW_t^{(2)}=\rho dt$. The remaining parameters -- $\mu$, $\theta$, $\kappa$, and $\sigma$ -- can be interpreted as the drift\index{Heston model!drift}, the long-run variance\index{Heston model!long-run variance}, the rate of mean reversion\index{Heston model!rate of mean reversion} to the long-run variance, and the volatility of variance\index{Heston model!volatility of variance} (often called the {\em vol of vol}\index{Heston model!vol of vol}), respectively. Sample paths and volatilities of the GBM and the Heston spot price process are plotted in Figure \ref{fig-STF2hes01}.

By setting $x_t = \log(S_t/S_0) - \mu t$, we can express the Heston model (\ref{eq:heston1})-(\ref{eq:volprocess}) in terms of the centered (log-)return $x_t$ and $v_t$. The process is then characterized by the transition probability $P_t(x,v\,|\,v_0)$ to have (log-)return $x$ and variance $v$ at time $t$ given the initial return $x=0$ and variance $v_0$ at time $t=0$. 
The time evolution of $P_t(x,v\,|\,v_0)$ is governed by the following Fokker-Planck (or forward Kolmogorov) equation:

\begin{eqnarray}\label{FP}
	\frac{\partial}{\partial t}P & = &
	     \kappa\frac{\partial}{\partial v}\left\{(v-\theta)P\right\}
	     + \frac12\frac{\partial}{\partial x}(vP) + \nonumber \\
    && + \, \rho\sigma\frac{\partial^2}{\partial x\,\partial v}(vP)
	     + \frac12\frac{\partial^2}{\partial x^2}(vP)  
	     + \frac{\sigma^2}{2}\frac{\partial^2}{\partial v^2}(vP).  
\end{eqnarray}

Solving this equation yields the following semi-analytical formula for the density of centered returns $x$, given a time lag $t$ of the price changes 
\cite{dra:yak:02}:\index{Heston model!density}
\begin{equation} \label{eq:HestonPfinal}
   P_t(x) = \frac{1}{2\pi}\int_{-\infty}^{+\infty} e^{i\xi x + F_t(\xi)} d\xi,
\end{equation}
with
\begin{eqnarray}
  & F_t(\xi) 
   = \frac{\kappa\theta}{\sigma^2}\, \gamma t - \frac{2\kappa\theta}{\sigma^2}
  	\log\left(\cosh\frac{\Omega t}{2} + 
    \frac{\Omega^2 -\gamma^2 + 2\kappa\gamma}{2\kappa\Omega}
    \sinh\frac{\Omega t}{2}\right), \nonumber \\
  & \gamma = \kappa + i\rho\sigma \xi, \quad \mbox{and} \quad 
  \Omega  = \sqrt{\gamma^2 + \sigma^2(\xi^2-i\xi)}. \nonumber 
\end{eqnarray}

Somewhat surprisingly, the introduction of SV does not change the properties of the spot price process in a way that could be noticed just by a visual inspection of its realizations, see Figure \ref{fig-STF2hes01} where sample paths of a GBM and the spot process (\ref{eq:heston1}) in the Heston model are plotted.\index{Heston model!simulation} To make the comparison more objective both trajectories were obtained with the same set of random numbers. 

In both cases the initial spot rate $S_0=4$ and the domestic and foreign interest rates are $r_d=5\%$ and $r_f=3\%$, respectively, yielding a drift of $\mu=r_d-r_f=2\%$. The volatility in the GBM is constant $\sqrt{v_t}=\sqrt{4\%}=20\%$, while in the Heston model it is driven by the mean-reverting process (\ref{eq:volprocess}) with the initial variance $v_0=4\%$, the long-run variance $\theta=4\%$, the speed of mean reversion $\kappa=2$, and the vol of vol $\sigma=30\%$. The correlation is set to $\rho=-0.05$.

\begin{figure}[tbp]
\begin{center}
\infig{1.4}{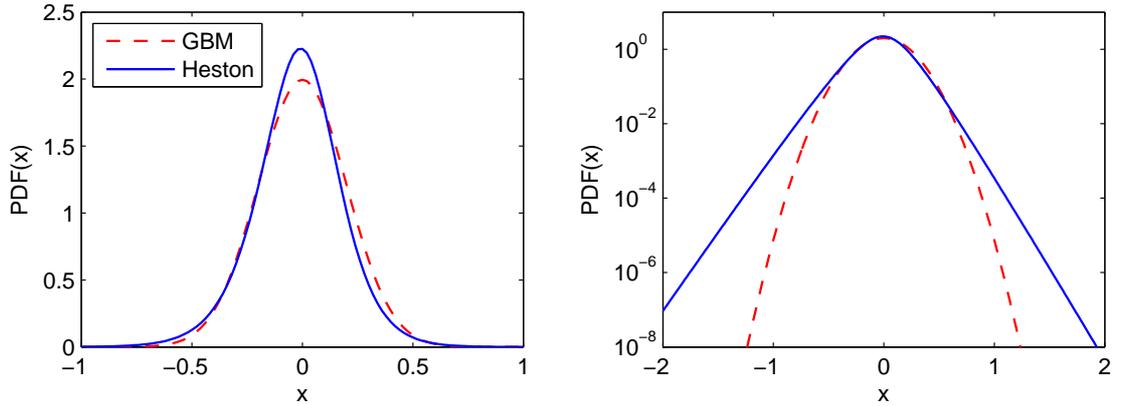}
\caption{The marginal pdfs in the Black-Scholes (GBM) and Heston models for the same set of parameters as in Figure \protect\ref{fig-STF2hes01} (\textit{left panel}). The tails of the Heston marginal pdfs are exponential, which is clearly visible in the log-linear scale (\textit{right panel}).}
\label{fig-STF2hes02}
\end{center}
\clink[m]{STFhes02}
\end{figure}

A closer inspection of the Heston model does, however, reveal some important differences with respect to GBM. For instance, the probability density functions (pdfs) of (log-)returns have heavier tails -- exponential compared to Gaussian, see Figure \ref{fig-STF2hes02}. In this respect they are similar to hyperbolic distributions \cite[see also Chapter 1]{wer:04}, i.e.\ in the log-linear scale they resemble hyperbolas, rather than parabolas of the Gaussian pdfs. 

\xsection{Option pricing}{hes_option}{}

Consider the value function of a general contingent claim $U(t,v,S)$ paying $g(S)=U(T,v,S)$ at time $T$. We want to replicate it with a self-financing portfolio. Due to the fact that in the Heston model we have two sources of uncertainty (the Wiener processes $W^{(1)}$ and $W^{(2)}$) the portfolio must include the possibility to trade in the money market, the underlying and another derivative security with value function $V(t,v,S)$. 

We start with an initial wealth $X_0$ which evolves according to: 
\begin{equation}
dX=\Delta\,dS+\Gamma\,dV+r_d(X-\Gamma V)\,dt-(r_d-r_f)\Delta S\,dt,
\end{equation}
where $\Delta$ is the number of units of the underlying held at time $t$ and $\Gamma$ is the number of derivative securities $V$ held at time $t$. 

The goal is to find $\Delta$ and $\Gamma$ such that $X_t=U(t,v_t,S_t)$ for all $t\in[0,T]$. The standard approach to achieve this is to compare the differentials of $U$ and $X$ obtained via It\^{o}'s formula. After some algebra we arrive at the partial differential equation (PDE) which $U$ must satisfy \citeaffixed{hak:wys:02}{for details on the derivation in the foreign exchange setting see}:
\begin{eqnarray}\label{eq:generalpde}
\frac{1}{2}vS^2 \frac{\partial^2 U}{\partial S^2} 
   + \rho\sigma vS \frac{\partial^2 U}{\partial S \partial v} 
   + \frac{1}{2}\sigma^2 v \frac{\partial^2 U}{\partial v^2}
   + (r_d-r_f)S \frac{\partial U}{\partial S}+ & & \nonumber\\
+ \Big\{\kappa(\theta-v)-\lambda(t,v,S) \Big\} \frac{\partial U}{\partial v} 
  - r_dU + \frac{\partial U}{\partial t} & = & 0.
\end{eqnarray}

The term $\lambda(t,v,S)$ is called the {\em market price of volatility risk}.\index{market price of volatility risk} \citeasnoun{hes:93} assumed it to be linear in the instantaneous variance $v_t$, i.e. $\lambda(t,v,S)=\lambda v_t$, in order to retain the form of the equation under the transformation from the statistical (or risky) measure to the risk-neutral measure.

\xsubsection{European vanilla FX option prices and Greeks}{hes_vanilla}{}

We can solve (\ref{eq:generalpde}) by specifying appropriate boundary conditions. For a European vanilla FX option these are:
\begin{eqnarray}
U(T,v,S) & = & \max\{\phi(S-K),0\},\\
U(t,v,0) & = & \frac{1-\phi}{2}Ke^{-r_d\tau},\\
\frac{\partial U}{\partial S}(t,v,\infty) & = & \frac{1+\phi}{2}e^{-r_f\tau},
\end{eqnarray}
\begin{eqnarray}
r_dU(t,0,S) & = & (r_d-r_f)S \frac{\partial U}{\partial S}(t,0,S)+ \nonumber\\
  & & + \, \kappa\theta \frac{\partial U}{\partial v}(t,0,S) 
  + \frac{\partial U}{\partial t}(t,0,S),\\
U(t,\infty,S) & = & \begin{cases} Se^{-r_f\tau}, \quad \mbox{for} ~~\phi=+1,\\
                    Ke^{-r_d\tau}, \quad \mbox{for} ~~\phi=-1, \end{cases}
\end{eqnarray}
where $\phi=\pm 1$ for call and put options, respectively. The strike $K$ is in units of the domestic currency and $\tau=T-t$ is the time to maturity (i.e. $T$ is the expiration time in years and $t$ is the current time).

\citeasnoun{hes:93} solved the PDE analytically using the method of characteristic functions. For European vanilla FX options the price is given by: 
\begin{eqnarray}\label{eq:hestonvanilla}
h(\tau) &=& \mbox{HestonVanilla}
	(\kappa,\theta,\sigma,\rho,\lambda,r_d,r_f,v_t,S_t,K,\tau,\phi)\nonumber\\
	& =& \phi \big\{ e^{-r_f\tau}S_tP_+(\phi)-Ke^{-r_d\tau}P_-(\phi) \big\},
\end{eqnarray}
where $u_{1,2}=\pm\frac{1}{2}$, $b_1=\kappa+\lambda-\sigma\rho$,
$b_2=\kappa+\lambda$, and
\begin{eqnarray}
d_j & = & \sqrt{(\rho\sigma\varphi i -b_j)^2-\sigma^2(2u_j\varphi i-\varphi^2)},\\ 
g_j & = & \frac{b_j-\rho\sigma\varphi i +d_j}{b_j-\rho\sigma\varphi i -d_j}, \label{eqn:g_j}\\
C_j(\tau,\varphi)&=&(r_d-r_f)\varphi i\tau + \label{eqn:C_j} \\
  & &+ \, \frac{\kappa\theta}{\sigma^2}\left\{(b_j-\rho\sigma\varphi i + 
     d_j)\tau-2\log\left(\frac{1-g_je^{d_j\tau}}{1-g_j}\right)\right\},
     \nonumber \\
D_j(\tau,\varphi)&=& \frac{b_j-\rho\sigma\varphi i 
  + d_j}{\sigma^2}\left(\frac{1-e^{d_j\tau}}{1-g_je^{d_j\tau}}\right),  \label{eqn:D_j}
\end{eqnarray}
\begin{eqnarray}  
f_j(x,v_t,\tau,\varphi)&=&\exp\{C_j(\tau,\varphi)+D_j(\tau,\varphi)v_t+i\varphi x\}, \label{eqn:f_j} \\
P_j(x,v_t,\tau,y)&=&\frac{1}{2}+\frac{1}{\pi}\int_0^{\infty}\Re\left\{\frac{e^{-i\varphi 
    y}f_j(x,v_t,\tau,\varphi)}{i\varphi}\right\}\,d\varphi.\label{eqn:P_j} 
\end{eqnarray}
Note that the functions $P_j$ are the cumulative distribution functions (in the variable $y=\log K$) of the log-spot price after time $\tau =T-t$ starting at $x=\log S_t$ for some drift $\mu$. Finally:
\begin{eqnarray}
P_+(\phi)&=&\frac{1-\phi}{2}+\phi P_1(x,v_t,\tau,y),\label{eqn:Pplus}\\
P_-(\phi)&=&\frac{1-\phi}{2}+\phi P_2(x,v_t,\tau,y).\label{eqn:Pminus}
\end{eqnarray}

The Greeks\index{Heston model!greeks} can be evaluated by taking the appropriate derivatives or by exploiting  homogeneity properties of financial markets \cite{rei:wys:01}. In the Heston model the \emph{spot delta} and the so-called \emph{dual delta} are given by:
\begin{equation}\label{eq:hes-delta}
\Delta=\frac{\partial h(t)}{\partial S_t}=\phi e^{-r_f\tau}P_+(\phi) \quad \mbox{and} \quad
\frac{\partial h(t)}{\partial K}=-\phi e^{-r_d\tau}P_-(\phi),
\end{equation}
respectively. \emph{Gamma}, which measures the sensitivity of delta to the underlying has the form:
\begin{equation}
\Gamma = \frac{\partial \Delta}{\partial S_t}=\frac{ e^{-r_f\tau}}{S_t}p_1(\log S_t, v_t, \tau, \log K),
\end{equation}
where
\begin{equation}
p_j(x,v,\tau,y)=\frac{1}{\pi}\int_0^{\infty}\Re\left\{e^{-i\varphi     y}f_j(x,v,\tau,\varphi)\right\}\,d\varphi, \quad j=1,2, \label{eqn:p_j}
\end{equation}
are the densities corresponding to the cumulative distribution functions $P_j$ (\ref{eqn:P_j}).

The time sensitivity parameter $theta = \partial h(t) / \partial t$ can be easily computed from (\ref{eq:generalpde}), while the formulas for \emph{rho} are the following: 
\begin{equation}
\frac{\partial h(t)}{\partial r_d}=\phi Ke^{-r_d\tau}\tau P_-(\phi), \qquad
\frac{\partial h(t)}{\partial r_f}=-\phi S_te^{-r_f\tau}\tau P_+(\phi).
\end{equation}
Note, that in the foreign exchange setting there are two \emph{rho}'s -- one is a derivative of the option price with respect to the domestic interest rate and the other is a derivative  with respect to the foreign interest rate.

The notions of \emph{vega} and \emph{volga} usually refer to the first and second derivative with respect to volatility. In the Heston model we use them for the first and second derivative with respect to the {\em initial variance}:
\begin{eqnarray}
\frac{\partial h(t)}{\partial v_t}&=&e^{-r_f\tau}S_t\frac{\partial }{\partial v_t}P_1(\log S_t,v_t,\tau,\log K)- \nonumber\\
& &- \, Ke^{-r_d\tau}\frac{\partial }{\partial v_t}P_2(\log S_t,v_t,\tau,\log K),\\
\frac{\partial^2 h(t)}{\partial v_t^2}&=&e^{-r_f\tau}S_t\frac{\partial^2 }{\partial v_t^2}P_1(\log S_t,v_t,\tau,\log K) - \nonumber\\
& &- \, Ke^{-r_d\tau}\frac{\partial^2 }{\partial v_t^2}P_2(\log S_t,v_t,\tau,\log K),
\end{eqnarray}
where
\begin{eqnarray}
\frac{\partial }{\partial v_t}P_j(x,v_t,\tau,y)&=&\frac{1}{\pi}    
  \int_0^{\infty}\Re\left[\frac{D(\tau,\varphi)e^{-i\varphi 
  y}f_j(x,v_t,\tau,\varphi)}{i\varphi}\right]\,d\varphi,\\
\frac{\partial^2 }{\partial v_t^2}P_j(x,v_t,\tau,y)&=&\frac{1}{\pi}  
  \int_0^{\infty}\Re\left[\frac{D^2(\tau,\varphi)e^{-i\varphi 
  y}f_j(x,v_t,\tau,\varphi)}{i\varphi}\right]\,d\varphi.
\end{eqnarray}

\xsubsection{Computational issues}{hes_compstat}{}

Heston's solution is semi-analytical. Formulas (\ref{eqn:Pplus}-\ref{eqn:Pminus}) require to integrate functions $f_j$, which are typically of oscillatory nature. \citeasnoun{hak:wys:02} propose to perform the integration in (\ref{eqn:P_j}) with the Gauss-Laguerre quadrature using $100$ for $\infty$ and $100$ abscissas. \citeasnoun{jac:kah:05} suggest using the Gauss-Lobatto quadrature (e.g.\ Matlab's \emph{quadl.m} function) and transform the original integral boundaries $[0,+\infty)$ to the finite interval $[0,1]$. 

As a number of authors have recently reported \cite{alb:may:sch:tis:06,gat:06,jac:kah:05}, the real problem starts when the functions $f_j$ are evaluated as part of the quadrature scheme. In particular, the calculation of the complex logarithm in eqn.\ (\ref{eqn:C_j}) is prone to numerical instabilities. It turns out that taking the principal value of the logarithm causes $C_j$ to jump discontinuously each time the imaginary part of the argument of the logarithm crosses the negative real axis. 
One solution is to keep track of the winding number in the integration (\ref{eqn:P_j}), but is difficult to implement because standard numerical integration routines cannot be used. \citeasnoun{jac:kah:05} provide a practical solution to this problem.

A different approach was taken by \citeasnoun{alb:may:sch:tis:06}, see also \citeasnoun{gat:06}. They proposed to make a simple transformation when computing the characteristic function and proved that the numerical stability of the resulting formulas is guaranteed under the full dimensional and unrestricted parameter space. Namely, their idea is to switch from $g_j$ in (\ref{eqn:g_j}) to 
\begin{equation}\label{eqn:tilde:g_j}
\tilde{g}_j = \frac{1}{g_j} = \frac{b_j-\rho\sigma\varphi i - d_j}{b_j-\rho\sigma\varphi i + d_j},
\end{equation}
which leads to new formulas for $C_j$ and $D_j$:
\begin{eqnarray}
{C}_j(\tau,\varphi)&=&(r_d-r_f)\varphi i\tau + \label{eqn:tilde:C_j}\\
  & &+ \, \frac{\kappa\theta}{\sigma^2}\left\{(b_j-\rho\sigma\varphi i - 
     d_j)\tau - 2\log\left(\frac{1 - \tilde{g}_je^{-d_j\tau}}{1 - \tilde{g}_j}\right)\right\}, \nonumber\\
{D}_j(\tau,\varphi)&=& \frac{b_j-\rho\sigma\varphi i 
  - d_j}{\sigma^2}\left(\frac{1-e^{-d_j\tau}}{1-\tilde{g}_je^{-d_j\tau}}\right), \label{eqn:tilde:D_j}
\end{eqnarray}
in (\ref{eqn:f_j}). Note, that the only differences between eqns.\ (\ref{eqn:g_j})-(\ref{eqn:D_j}) and (\ref{eqn:tilde:g_j})-(\ref{eqn:tilde:D_j}), respectively, are the flipped minus and plus signs in front of the $d_j$'s.

The  mispricings resulting from using (\ref{eqn:g_j})-(\ref{eqn:D_j}) are not that obvious to notice. In fact, if we price and backtest on short or middle term maturities only, we might not detect the problem at all. However, the deviations can become extreme for longer maturities \citeaffixed{alb:may:sch:tis:06}{typically above 3-5 years; the exact threshold is parameter dependent, see}.

Apart from the above semi-analytical solution for vanilla options, alternative approaches can be utilized. These include the general Fast Fourier Transform (FFT)\index{Fast Fourier Transform (FFT)} approach of \citeasnoun{car:mad:99}, finite differences, finite element methods and  Monte Carlo simulations. The FFT-based pricing technique is discussed in Section \ref{hes_fft}. As for the other methods, finite differences must be used with care since high precision is required to invert scarce matrices. The Crank-Nicholson, ADI (Alternate Direction Implicit), and Hopscotch schemes can be used, however, ADI is not suitable to handle nonzero correlation. Boundary conditions must be set appropriately, for details see \citeasnoun{klu:02}. On the other hand, finite element methods can be applied to price both the vanillas and exotics, as explained for example in \citeasnoun{ape:win:wys:02}. 

Finally, the Monte Carlo approach requires attention as the simple Euler discretization of the CIR process (\ref{eq:volprocess}) may give rise to a negative variance.\index{Heston model!simulation}
To deal with this problem, practitioners generally either adopt (i) the absorbing $v_t = \max(v_t,0)$, or (ii) the reflecting principle $v_t = \max(v_t,-v_t)$. More elegant, but computationally more demanding solutions include sampling from the exact transition law \cite{gla:04} or application of the quadratic-exponential (QE) scheme \cite{and:08}. For a recent survey see the latter reference, where several new algorithms for time-discretization and Monte Carlo simulation of Heston-type SV models are introduced and compared in a thorough simulation study. Both the absorbing/reflecting patches and the QE scheme are implemented in the \emph{simHeston.m} function used to plot Figure \ref{fig-STF2hes01}.

\xsubsection{Behavior of the variance process and the Feller condition}{hes_calibration_feller}{}
\index{Heston model!Feller condition}

The CIR process for the variance, as defined by (\ref{eq:volprocess}), always remains non-negative. This is an important property which, for instance, is not satisfied by the Ornstein-Uhlenbeck process. However, ideally one would like a variance process which is strictly positive, because otherwise the spot price process degenerates to a deterministic function for the time the variance stays at zero. As it turns out, the CIR process remains strictly positive under the condition that
\begin{eqnarray}
\label{eq_feller_condition}
\alpha: = \frac{4\kappa\theta}{\sigma^2} \ge 2,
\end{eqnarray}
which is often referred to as the \emph{Feller condition}. We call $\alpha$ the \emph{dimensionality} of the corresponding Bessel process (see below). If the condition is not satisfied, i.e.\ for $0 < \alpha < 2$, the CIR process will visit $0$ recurrently but will not stay at zero, i.e.\ the $0$-boundary is strongly reflecting.

Unfortunately, when calibrating the Heston model to market option prices it is not uncommon for the parameters to violate the Feller condition (\ref{eq_feller_condition}). This is not a complete disaster, as the variance process can only hit zero for an infinitesimally small amount of time, but it is still worrying as very low levels of volatility (e.g.\ say below 1\%) are repeatedly reached for short amounts of time and that is not something observed in the market.

Besides being important from a modeling point of view, the Feller condition also plays a role in computational accuracy. For Monte Carlo simulations special care has to be taken so that the simulated paths do not go below zero if (\ref{eq_feller_condition}) is not satisfied. 
On the PDE side, the Feller condition determines whether the zero-variance boundary is in- or out-flowing, that is to say whether the convection vector at the boundary points in- or outwards.
To see this, we need to write the log-spot transformed Heston PDE in convection-diffusion form
\begin{eqnarray}
  \frac{\partial}{\partial t} U = \divergence(A\grad U)
   - \divergence(U b) + f ,
\end{eqnarray}
and obtain
\begin{eqnarray}
 b(x,v) = v
                \begin{pmatrix}
                        \frac{1}{2} \\
                        \kappa+\lambda
                \end{pmatrix}+
                \begin{pmatrix}
                        \frac{1}{2}\rho\sigma + r_f-r_d \\
                        \frac{1}{2}\sigma^2 -\kappa\theta
                \end{pmatrix},  
\end{eqnarray}
which is out-flowing at the $v=0$ boundary if
\begin{eqnarray}
    \frac{1}{2}\sigma^2 -\kappa\theta < 0 ,
\end{eqnarray}
in which case the Feller condition (\ref{eq_feller_condition}) is satisfied.

Having introduced and discussed the importance of this condition, we now give an overview of how it is derived; we refer the interested reader to Chapter 6.3 in \citeasnoun{jea:yor:che:09} for a more thorough treatment.
The main idea is to relate the CIR process to the family of squared Bessel processes which have well known properties. We call $X_t$ an $\alpha$-dimensional squared Bessel process and denote it by $\Bessel^2(\alpha)$ if it follows the dynamics of
\begin{eqnarray}
   \dd X_t = \alpha \dd t + 2 \sqrt{X_t^+} \dd W_t ,
\end{eqnarray}
with $X_t^+:=\max\{0,X_t\}$. This definition makes sense for any real valued $\alpha$. However, in the case of integer valued $\alpha$ we have an interesting interpretation: a $\Bessel^2(\alpha)$ process $X_t$, with $X_0=0$, follows the same dynamics as the squared distance to the origin of an $\alpha$-dimensional Brownian motion, i.e.\ 
\begin{eqnarray}
   X_t = \sum_{i=1}^{\alpha} B_t^{(i)} ,
\end{eqnarray}
with $B^{(i)}$ being independent Brownian motions. From this we can  conclude that for $\alpha=2$, $X_t$ will never reach zero and, using the stochastic comparison theorem \cite{rog:wil:00}, this property remains true for any $\alpha\ge 2$. Similarly for $0<\alpha\leq 1$ the value zero will be repeatedly hit (for $\alpha=1$ this happens as often as a one-dimensional Brownian motion crosses zero). 

The stochastic comparison theorem also immediately tells us that $\Bessel^2(\alpha)$ processes are non-negative for non-negative $\alpha$: for $\alpha=0$ we get the trivial solution $X_t=0$ and increasing the drift term to $\alpha>0$ cannot make the paths any smaller.
For a proof of the above statements we refer to Chapter V.48 in \citeasnoun{rog:wil:00}, from which we also state the following additional properties.
If $X_t$ is an $\alpha$-dimensional squared Bessel process $\Bessel^2(\alpha)$ and $X_0\ge 0$ then it is always non-negative and:
 \begin{enumerate}
  \item for $0<\alpha<2$ the process hits zero and this is recurrent, but
	the time spent at zero is zero,
  \item for $\alpha=2$ the process is strictly positive but gets arbitrarily close to $0$ and $\infty$,
  \item for $\alpha>2$ the process is strictly positive and tends to infinity as time approaches infinity.
 \end{enumerate}
To translate these properties to the class of CIR processes, we only need to apply the following space-time transformation to the squared Bessel process. Define  $\dd Y_t := \e^{-\kappa t} X_{\beta (\e^{\kappa t}-1)}$. Then $Y_t$ follows the dynamics of
\begin{eqnarray}
   \dd Y_t = \kappa(\alpha \beta - Y_t) \dd t + 2 \sqrt{\kappa\beta Y_t} \dd W_t,
\end{eqnarray}
which is the same as the dynamics of the CIR process (\ref{eq:volprocess}) if we set
$\beta=\tfrac{\sigma^2}{4\kappa}$ and 
$\alpha=\tfrac{\theta}{\beta}=\tfrac{4\kappa\theta}{\sigma^2}$.

\xsubsection{Option pricing with FFT}{hes_fft}{}
\index{Fast Fourier Transform (FFT)!option pricing}

In this section, we briefly describe the numerical option pricing approach of \citeasnoun{car:mad:99}, which utilizes the characteristic function (cf) of the underlying instrument's price process. The basic idea of the method is to compute the price by Fourier inversion given an analytic expression for the cf of the option price. 

The rationale for using this approach is twofold. Firstly, the algorithm offers a speed advantage, including the possibility to calculate prices for a whole range of strikes in a single run. Secondly, the cf of the log-price is known and has a simple form for many models, while the pdf is often either unknown in closed-form or complicated from the numerical point of view. For instance, for the Heston model the cf takes the form \cite{hes:93,jac:kah:05}:\index{Heston model!characteristic function (cf)}
\begin{equation}
\E\{\exp(i\varphi\log S_T)\} = f_2(x,v_t,\tau,\varphi), \label{eqn:CF}
\end{equation}
where $f_2$ is given by (\ref{eqn:f_j}).

Let us now derive the formula for the price of a European vanilla call option. Derivation of the put option price follows the same lines, for details see \citeasnoun{lee:04} and \citeasnoun{sch:10}. Alternatively we can use the call-put parity for European vanilla FX options \citeaffixed{wys:06}{see e.g.}. Let $h^C(\tau;k)$ denote the price of the call option maturing in $\tau=T-t$ years with a strike of $K= \exp(k)$:
\begin{equation}
h^C(\tau;k)=\int_k^\infty e^{-rT} (e^s-e^k)q_T(s)ds,
\end{equation}
where $q_T$ is the risk-neutral density of $s_T= \log S_T$. The function $h^C(\tau;k)$ is not square-integrable \citeaffixed{rud:91}{see e.g.} because it converges to $S_0$ for $k \to -\infty$. Hence, consider a modified function $H^C(\tau;k)= e^{\alpha k} h^C(\tau;k)$, which is square-integrable for a suitable constant $\alpha >0$. 
A sufficient condition 
is given by:  
\begin{equation}\label{eqn:condition}
\E\{(S_T)^{\alpha +1}\} < \infty,
\end{equation}
which is equivalent to $\psi_T(0)$, i.e.\ the Fourier transform of $H^C(\tau;k)$, see (\ref{eqn:psi_T}) below, being finite.
In an empirical study \citeasnoun{sch:sim:tis:04} found that $\alpha=0.75$ leads to stable algorithms, i.e.\ the prices are well replicated for many model parameters. This value also fulfills condition (\ref{eqn:condition}) for the Heston model \cite{bor:det:hae:05}. Note, that for put options the condition is different: it is sufficient to choose $\alpha>0$ such that $E\{(S_T)^{-\alpha}\} < \infty$ \cite{lee:04}.

\begin{figure}[tbp]
\begin{center}
\infig{1.4}{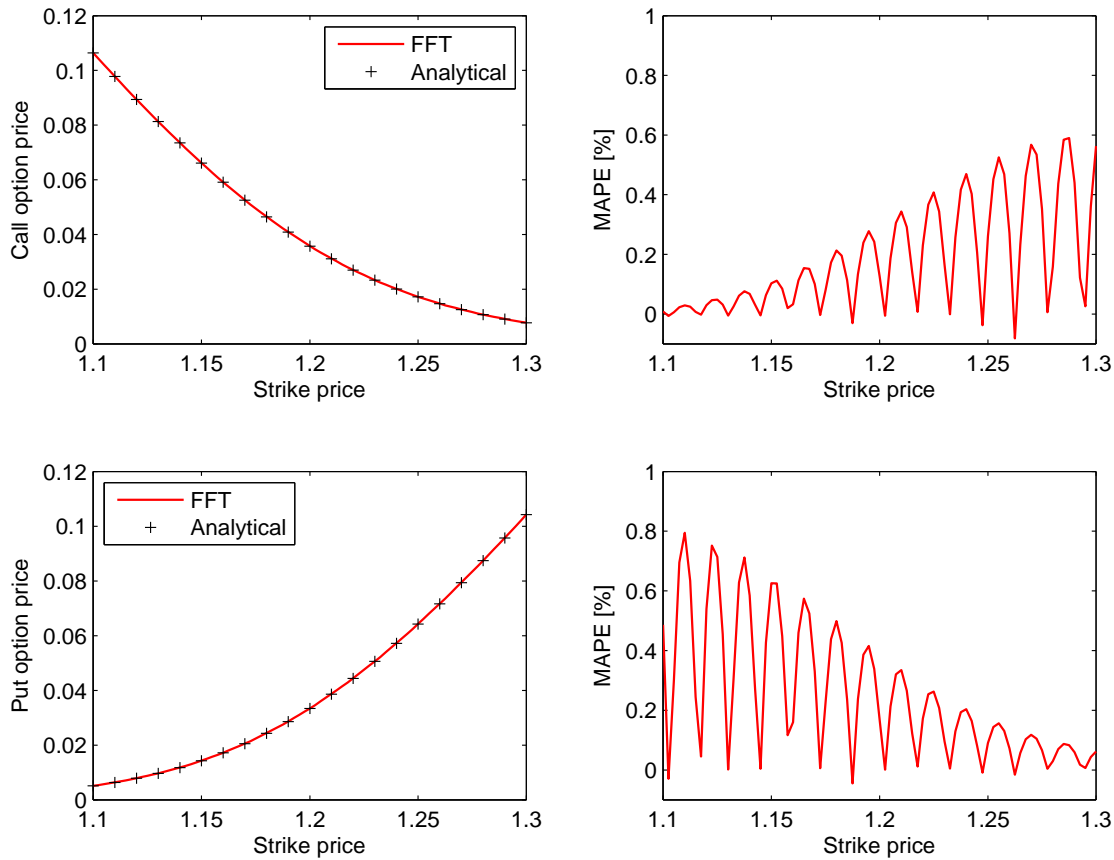}
\caption{European call (\emph{top left}) and put (\emph{bottom left}) FX option prices obtained using the FFT method and the `Analytical' formula (\ref{eq:hestonvanilla}) for a sample set of parameters. \textit{Right panels}: The percentage differences between the prices: $(\mbox{FFT}-\mbox{Analytical})/\mbox{Analytical} \times 100\%$.}
\label{fig-STF2hes03}
\end{center}
\clink[m]{STFhes03}
\end{figure}

Now, compute the Fourier transform of $H^C(\tau;k)$:\index{Fast Fourier Transform (FFT)}
\begin{eqnarray}
\psi_T(v) & = & \int_{- \infty}^\infty e^{ivk} H^C(\tau;k) dk \nonumber \\ 
& = &  \int_{-\infty}^\infty e^{ivk} \int_k^\infty e^{ \alpha k} e^{ - r T} (e^s - e^k ) q_T(s) ds dk \nonumber \\
& = & \int_{-\infty}^\infty e^{ - r T} q_T(s) \int_{-\infty }^s
\left\{e^{\alpha k+s} - e^{ ( \alpha +1)k} \right\} e^{i v k} dk ds \nonumber \\
& = & \int_{-\infty}^\infty e^{ - r T} q_T(s) \left\{ \frac{e^{(\alpha
+1+i v ) s} } { \alpha + i v} - \frac {e^{ (\alpha +1+ i v
) s}} { \alpha +1 +i v} \right\} ds \nonumber \\
& = & \frac{e^{-rT} f_2 \{v-(\alpha+1)i\}}{\alpha^2+\alpha-v^2+i
(2\alpha +1)v},\label{eqn:psi_T}
\end{eqnarray}
where $f_2$ is the cf of $q_T$, see (\ref{eqn:CF}). We get the option price in terms of $\psi_T$ using Fourier inversion:
\begin{equation}
h^C(\tau;k)=\frac{\exp (-\alpha k)}{\pi} \int_0^\infty e^{- i vk} \psi (v) dv.
\end{equation}
This integral can be numerically approximated as \cite{car:mad:99}:
\begin{equation}
h^C(\tau;k_u)\approx \frac{e^{-\alpha k_u}}{\pi} \sum_{j=1}^{N}
e^{-\frac{2\pi i}{N}(j-1)(u-1)} e^{i b v_j} \psi (v_j) \frac{\eta}{3} 
\{ 3+(-1)^j- \delta_{j-1}\},
\end{equation}
where $k_u = \frac{1}{\eta}\{-\pi + \frac{2\pi}{N}(u-1)\}$, $u=1,\ldots,N$, $\eta >0$ is the distance between the points of the integration grid, $v_j = \eta (j-1)$, $j=1,\ldots,N$, and $\delta$ is the Dirac function. 

As can be seen in Figure \ref{fig-STF2hes03} the differences between the FFT method and the (semi-)analytical formula (\ref{eq:hestonvanilla}) are relatively small. The differences result form the fact that the former method yields `exact' values only on the grid $k_u$. In order to preserve the speed of the FFT-based algorithm we use linear interpolation between the grid points. This approach, however, slightly overestimates the true prices since the option price is a convex function of the strike \cite{bor:det:hae:05}. It can be clearly seen that near the grid points the prices obtained by both methods coincide, while between the grid points the FFT-based algorithm generates higher prices than the analytical solution, see the right panels in Figure \ref{fig-STF2hes03}.

\xsection{Calibration}{hes_Calibration}{}

\xsubsection{Qualitative effects of changing the parameters}{hes_parameffects}{}

Before calibrating the model to market data we will show how changing the input parameters affects the shape of the fitted smile curve.\index{volatility smile} This analysis will help in reducing the dimensionality of the problem. In all plots of this subsection the solid blue curve with x's is the smile obtained for $v_0=0.01$, $\sigma = 0.2$, $\kappa = 1.5$, $\theta = 0.015$, and $\rho = 0.05$.

\begin{figure}[tbp]
\begin{center}
\infig{1.4}{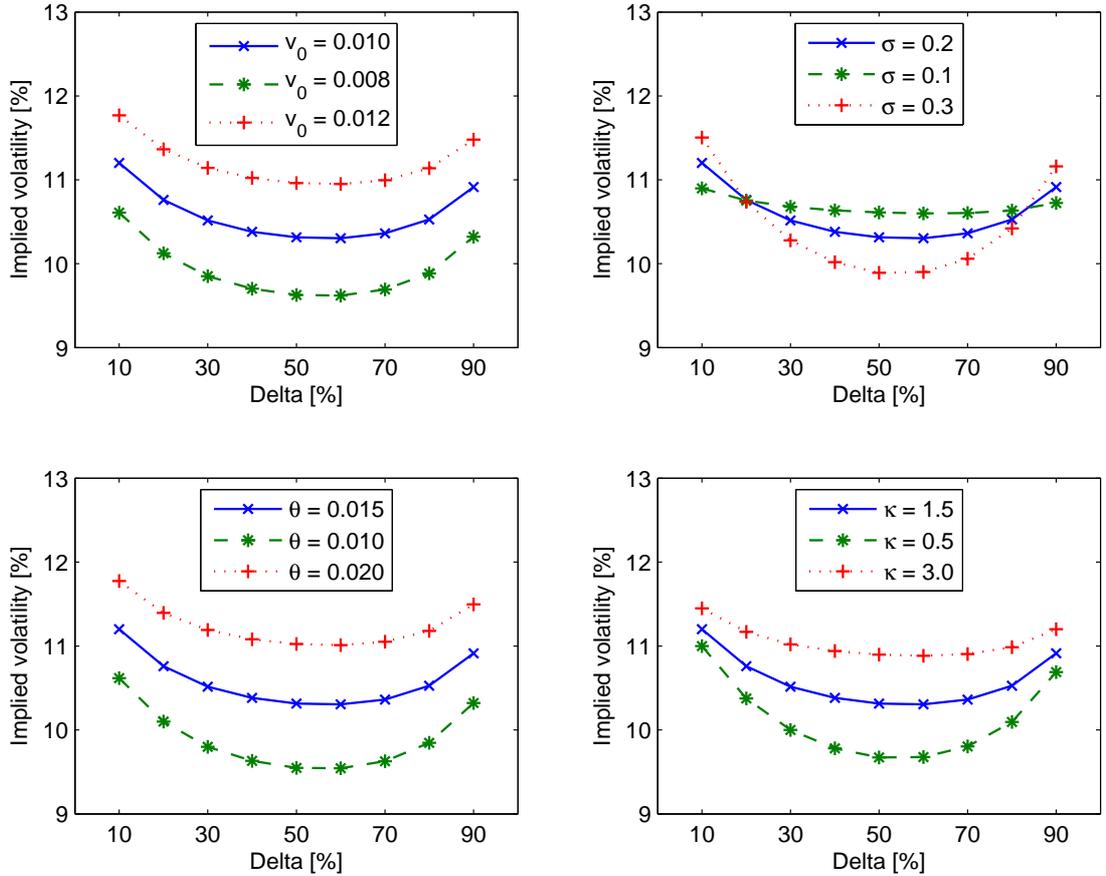}
\caption{The effects of changing the model parameters on the shape of the smile: initial variance $v_0$ (\textit{top left}), volatility of variance $\sigma$ (\textit{top right}), long-run variance $\theta$ (\textit{bottom left}), and mean reversion level $\kappa$ (\textit{bottom right}).}
\label{fig-STF2hes04}
\end{center}
\clink[m]{STFhes04}
\end{figure}

First, take a look at the initial variance\index{Heston model!initial variance} (top left panel in Figure \ref{fig-STF2hes04}). Apparently, changing $v_0$ allows for adjustments in the height of the smile curve. On the other hand, the volatility of variance (vol of vol)\index{Heston model!vol of vol} has a different impact on the smile. Increasing $\sigma$ increases the convexity of the fit, see the top right panel in Figure \ref{fig-STF2hes04}. In the limiting case, setting $\sigma$ equal to zero would produce a deterministic process for the variance and, hence, a volatility which does not admit any smile (a constant curve). 

The effects of changing the long-run variance $\theta$\index{Heston model!long-run variance} are similar to those observed by changing the initial variance, compare the left panels in Figure \ref{fig-STF2hes04}. It seems promising to choose the initial variance a priori, e.g. set $\sqrt{v_0}=$ implied at-the-money (ATM) volatility, and only let the long-run variance $\theta$ vary. In particular, a different initial variance for different maturities would be inconsistent. 

Changing the mean reversion $\kappa$\index{Heston model!rate of mean reversion} affects the ATM part more than the extreme wings of the smile curve. The low/high deltas ($\Delta$) remain almost unchanged while increasing the mean reversion lifts the center, see the bottom right panel in Figure \ref{fig-STF2hes04}. Moreover, the influence of $\kappa$ is often compensated by a stronger vol of vol $\sigma$. This suggests fixing mean reversion (at some level, say $\kappa=1.5$) and only calibrating the remaining three parameters. If the obtained parameters do not satisfy the Feller condition (\ref{eq_feller_condition}), it might be worthwhile to fix $\kappa$ at a different level, say $\tilde{\kappa}=3$, recalibrate the remaining parameters and check if the new estimates fulfill the condition and lead to a more realistic variance process.

\begin{figure}[tbp]
\begin{center}
\infig{1.4}{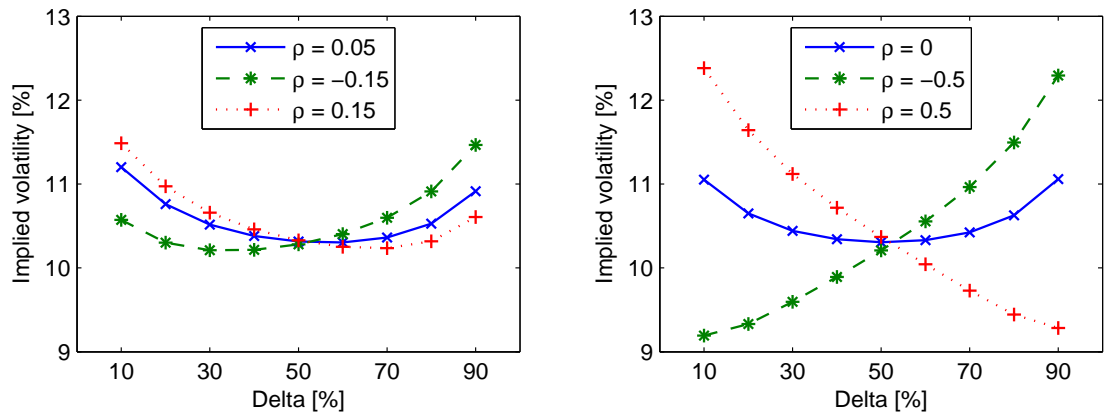}
\caption{The effects of changing the correlation on the shape of the smile.}
\label{fig-STF2hes05}
\end{center}
\clink[m]{STFhes05}
\end{figure}

Finally, let us look at the influence of correlation.\index{Heston model!correlation} The uncorrelated case produces a fit that looks like a symmetric smile curve centered at-the-money, see Figure \ref{fig-STF2hes05}. However, it is not exactly symmetric. Changing $\rho$ changes the degree of symmetry. In particular, positive correlation makes calls more expensive, negative correlation makes puts more expensive. 
Note, that the model yields a volatility skew, a typically observed volatility structure in equity markets, only when the correlation is set to a very high or low value

\xsubsection{The calibration scheme}{hes_calibration}{}
\index{Heston model!calibration}

Calibration of SV models can be done in two conceptually different ways. One is to look at the time series of historical data. Estimation methods such as Generalized, Simulated, and Efficient Methods of Moments (respectively GMM, SMM, and EMM), as well as Bayesian MCMC have been extensively applied. See \citeasnoun{bro:rui:04} for a review. In the Heston model we could also try to fit empirical distributions of returns to the marginal distributions specified in (\ref{eq:HestonPfinal}) via a minimization scheme. Unfortunately, all historical approaches have one common flaw: they do not allow for the estimation of the market price of volatility risk $\lambda(t,v,S)$.\index{market price of volatility risk} Observing only the underlying spot price and estimating SV models with this information will not yield correct prices for the derivatives. 

This leads us to the second estimation approach: instead of using the spot data we calibrate the model to the volatility smile\index{volatility smile} (i.e. prices of vanilla options). In this case we do not need to worry about estimating the market price of volatility risk as it is already embedded in the market smile. This means that we can set $\lambda=0$ by default and just determine the remaining parameters. 

As a preliminary step, we have to retrieve the strikes since the smile in FX markets is specified as a function of delta. Comparing the Black-Scholes type formulas (in the FX market setting we have to use the \citeasnoun{gar:koh:83} specification) for delta and the option premium yields the relation for the strikes $K_i$. From a computational point of view this stage requires only an inversion of the Gaussian distribution function.
Next, based on the findings of Section \ref{hes_parameffects}, we fix two parameters (initial variance $v_0$ and mean reversion $\kappa$) and fit the remaining three:  volatility of variance $\sigma$, long-run variance $\theta$, and correlation $\rho$ for a fixed time to maturity and a given vector of market Black-Scholes implied volatilities\index{implied volatility} $\{\hat{\sigma}_i\}_{i=1}^n$ for a given set of delta pillars $\{\Delta_i\}_{i=1}^n$.\index{delta pillar}

After fitting the parameters we compute the option prices in the Heston model using (\ref{eq:hestonvanilla}) and retrieve the corresponding Black-Scholes model implied volatilities $\{\sigma _i\}_{i=1}^n$ via a standard bisection method (function \emph{fzero.m} in Matlab uses a combination of bisection, secant, and inverse quadratic interpolation methods). The next step is to define an objective function, which we choose to be the Sum of Squared Errors (SSE):
\begin{eqnarray}
\mbox{SSE}(\kappa,\theta, \sigma, \rho, v_0) = \sum_{i=1}^n \{\hat{\sigma}_i-\sigma _i(\kappa,\theta, \sigma, \rho, v_0) \}^2.
\end{eqnarray}

We compare the volatilities because they are all of similar magnitude, in contrast to the prices which can range a few orders of magnitude for it-the-money (ITM) vs. out-of-the-money (OTM) options. In addition, one could introduce weights for all the summands to favor ATM or OTM fits. Finally we minimize over this objective function using a simplex search routine (\emph{fminsearch.m} in Matlab) to find the optimal set of parameters.

\xsubsection{Sample calibration results}{hes_calibration_results}{}
\index{Heston model!calibration}

\begin{figure}[tbp]
\begin{center}
\infig{1.4}{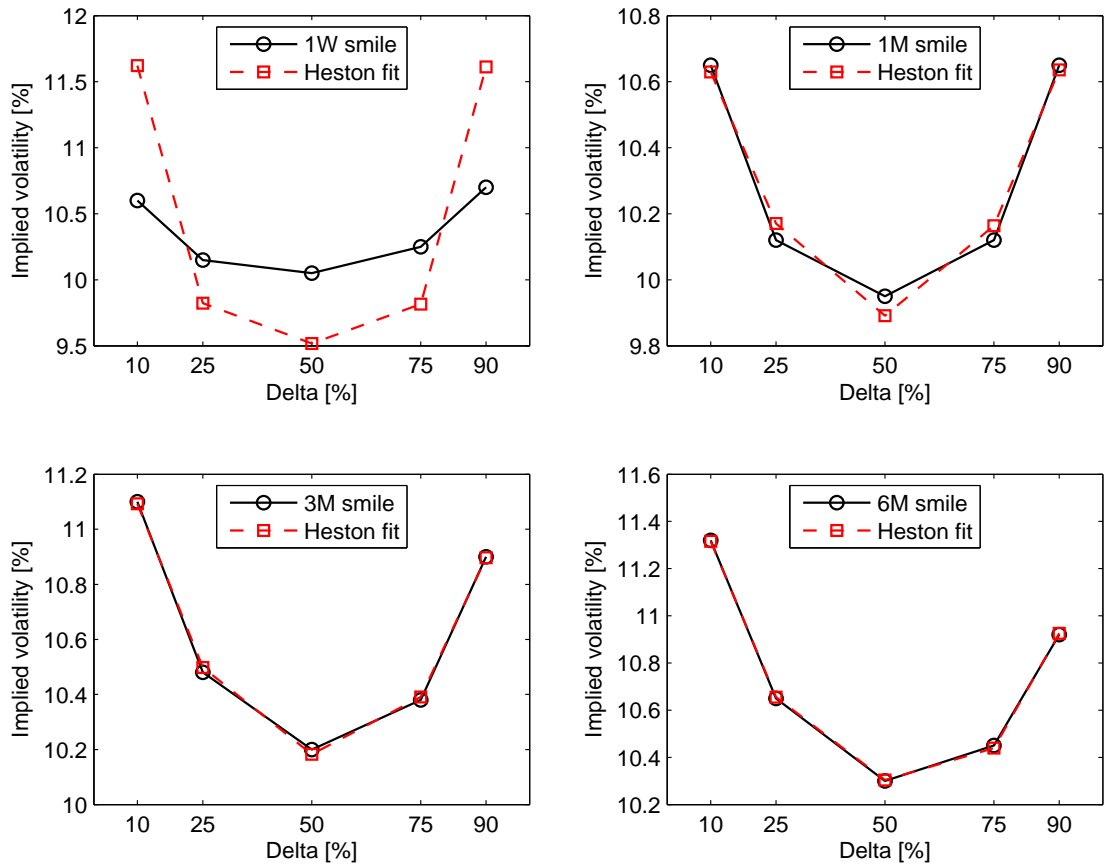}
\caption{The EUR/USD market smile on July 1, 2004 and the fit obtained with the Heston model for $\tau$ = 1 week (\textit{top left}), 1 month (\textit{top right}), 3 months (\textit{bottom left}), and 6 months (\textit{bottom right}).}
\label{fig-STF2hes06a}
\end{center}
\clink[m]{STFhes06}
\end{figure}

\begin{figure}[tbp]
\begin{center}
\infig{1.4}{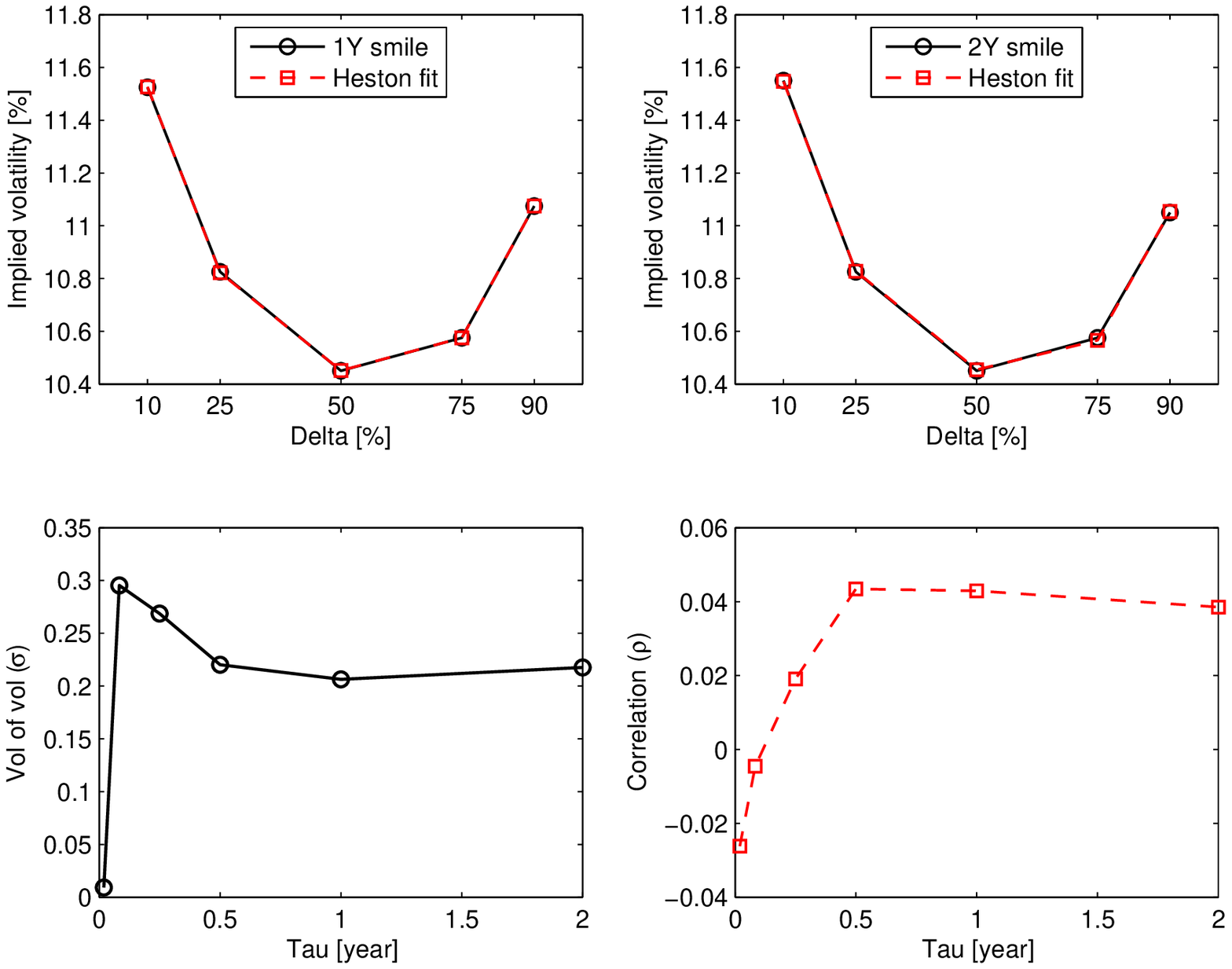}
\caption{The EUR/USD market smile on July 1, 2004 and the fit obtained with the Heston model for $\tau$ = 1 year (\textit{top left}) and 2 years (\textit{top right}). The term structure of the vol of vol and correlation visualizes the problem with fitting the smile in the short term (\textit{bottom panels}).}
\label{fig-STF2hes06b}
\end{center}
\clink[m]{STFhes06}
\end{figure}

We are now ready to calibrate the Heston model to market data. First, we take the EUR/USD volatility surface on July 1, 2004 and fit the parameters in the Heston model according to the calibration scheme discussed earlier. The results are shown in Figures \ref{fig-STF2hes06a}--\ref{fig-STF2hes06b}. Note, that the fit is very good for intermediate and long maturities (three months and more). Unfortunately, the Heston model does not perform satisfactorily for short maturities (see Section \ref{hes_JD} for a discussion of alternative approaches). Comparing with the fits in \citeasnoun{wer:wys:05} for the same data, the long maturity (2Y) fit is better. This is due to a more efficient optimization routine (Matlab 7.2 vs. XploRe 4.7) and utilization of the transformed formulas (\ref{eqn:tilde:g_j})-(\ref{eqn:tilde:D_j}) instead of the original ones (\ref{eqn:g_j})-(\ref{eqn:D_j}).

\begin{figure}[tbp]
\begin{center}
\infig{1.4}{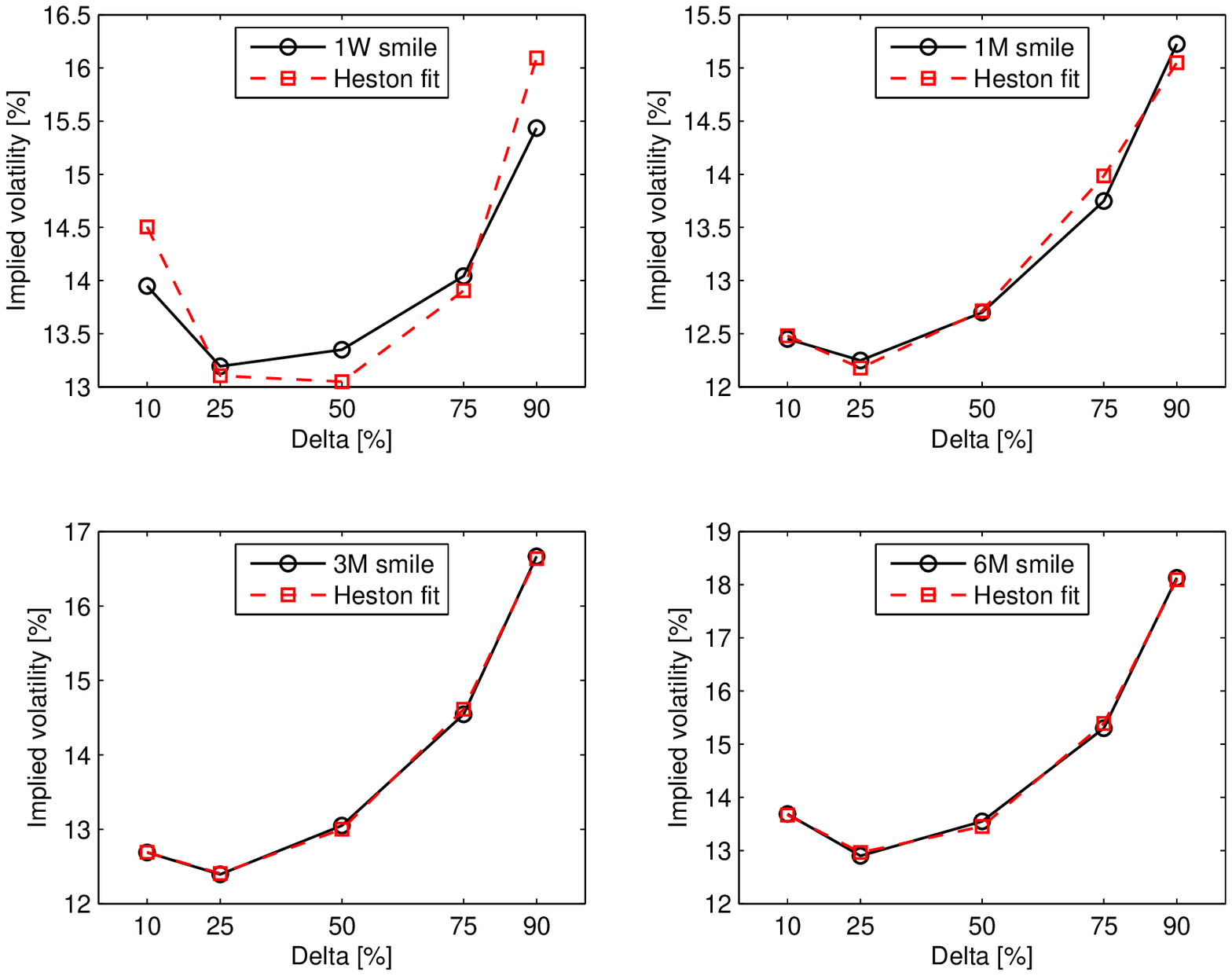}
\caption{The EUR/USD market smile on July 22, 2010 and the fit obtained with the Heston model for $\tau$ = 1 week (\textit{top left}), 1 month (\textit{top right}), 3 months (\textit{bottom left}), and 6 months (\textit{bottom right}).}
\label{fig-STF2hes07a}
\end{center}
\clink[m]{STFhes07}
\end{figure}

\begin{figure}[tbp]
\begin{center}
\infig{1.4}{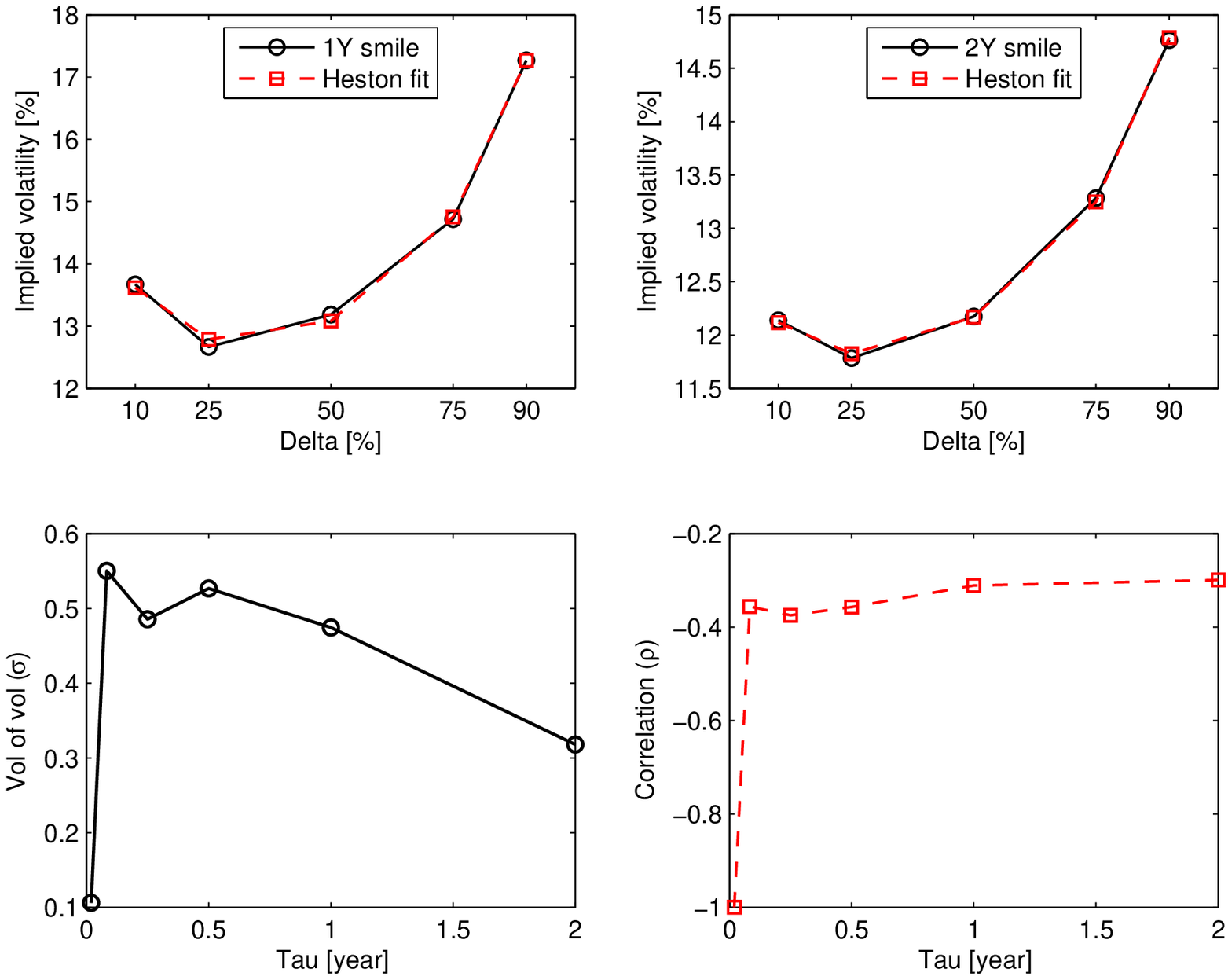}
\caption{The EUR/USD market smile on July 22, 2010 and the fit obtained with the Heston model for $\tau$ = 1 year (\textit{top left}) and 2 years (\textit{top right}). Again the term structure of the vol of vol and correlation visualizes the problem with fitting the smile in the short term (\textit{bottom panels}).}
\label{fig-STF2hes07b}
\end{center}
\clink[m]{STFhes07}
\end{figure}

Now we take a look at more recent data and calibrate the Heston model to the EUR/USD volatility surface on July 22, 2010. The results are shown in Figures \ref{fig-STF2hes07a}--\ref{fig-STF2hes07b}. Again the fit is very good for intermediate and long maturities, but unsatisfactory for maturities under three months. The term structure of the vol of vol and correlation visualizes the problem with fitting the smile in the short term for both datasets. The calibrated vol of vol is very low for the 1W smiles, then jumps to a higher level. The correlation behaves similarly, for the 1W smiles it is much lower than for the remaining maturities. In 2004 it additionally changes sign as the skew changes between short and longer-term tenors. Also note, that the more skewed\index{volatility smile}\index{volatility skew} smiles in 2010 require much higher (anti-)correlation ($-0.4 <\rho < -0.3$ for $\tau\ge 1\mbox{M}$) to obtain a decent fit than the more symmetric smiles in 2004 ($-0.01 <\rho < 0.05$ for $\tau\ge 1\mbox{M}$).

As these examples show, the Heston model can be successfully applied to modeling the volatility smile of vanilla FX options in the mid- to longer-term. There are essentially three parameters to fit, namely the long-run variance ($\theta$), which corresponds to the ATM level of the market smile, the vol of vol ($\sigma$), which corresponds to the convexity of the smile (in the market often quoted as \emph{butterflies}), and the correlation ($\rho$), which corresponds to the skew of the smile (quoted as \emph{risk reversals}). It is this direct link of the model parameters to the market that makes the Heston model so attractive to practitioners. 

The key application of the model is to calibrate it to vanilla options and afterward use it for pricing exotics \citeaffixed{wys:03}{like one-touch options, see} in either a finite difference grid or a Monte Carlo simulation. Surprisingly, the results often coincide with the \emph{traders' rule of thumb} pricing method \cite{wys:06}. This might also simply mean that a lot of traders use the same model. After all, it is a matter of belief which model reflects reality the best. Recent ideas are to take prices of one-touch options along with prices of vanilla options from the market and use this common input to calibrate the Heston model.

\xsection{Beyond the Heston model}{hes_beyond}{}

\xsubsection{Time-dependent parameters}{hes_time_dependent_param}{}

As we have seen in Section \ref{hes_calibration_results}, performing calibrations for different time slices of the volatility matrix produces different values of the parameters. This suggests a term structure of some parameters in the Heston model. Therefore, we need to generalize the CIR process (\ref{eq:volprocess}) to the case of time-dependent parameters, i.e.\ we consider the process:
\begin{equation}
dv_t=\kappa(t)\{\theta(t)-v_t\}\,dt+\sigma(t)\sqrt{v_t}\,dW_t,
\label{eq:CIRtermallprocess}
\end{equation}
for some nonnegative deterministic parameter functions $\sigma(t)$, $\kappa(t)$, and $\theta(t)$. The formula for the mean turns out to be:
\begin{equation}
\mbox{E}(v_t)= g(t)=v_0e^{-K(t)}+\int_0^t\kappa(s)\theta(s)e^{K(s)-K(t)}\,ds,
\end{equation}
with $K(t)= \int_0^t\kappa(s)\,ds$. The result for the second moment is:
\begin{eqnarray}
\mbox{E}(v_t^2)=v_0^2e^{-2K(t)}+\int_0^t \{2\kappa(s)\theta(s)+\sigma^2(s)\} g(s)e^{2K(s)-2K(t)}\,ds,
\end{eqnarray}
and hence for the variance (after some algebra):
\begin{eqnarray}
\mbox{Var}(v_t)=\int_0^t\sigma^2(s)g(s)e^{2K(s)-2K(t)}\,ds.
\end{eqnarray}

The formula for the variance allows us to compute forward volatilities of variance explicitly. Assuming known values $\sigma_{T_1}$ and $\sigma_{T_2}$ for times $0<T_1<T_2$, we want to determine the forward volatility of variance $\sigma_{T_1,T_2}$ which matches the corresponding variances, i.e.\ 
\begin{eqnarray}
& &\int_0^{T_2}\sigma_{T_2}^2g(s)e^{2\kappa(s-T_2)}\,ds =\\
& & \qquad =\int_0^{T_1}\sigma_{T_1}^2g(s)e^{2\kappa(s-T_2)}\,ds
+\int_{T_1}^{T_2}\sigma_{T_1,T_2}^2g(s)e^{2\kappa(s-T_2)}\,ds.  \nonumber
\end{eqnarray}
The resulting forward volatility of variance is thus:
\begin{equation}
\sigma_{T_1,T_2}^2=\frac{\sigma_{T_2}^2H(T_2)-\sigma_{T_1}^2H(T_1)}{H(T_2)-H(T_1)},
\end{equation}
where
\begin{equation}
	H(t)=\int_0^tg(s)e^{2\kappa s}\,ds
	=\frac{\theta}{2\kappa}e^{2\kappa t}+\frac{1}{\kappa}(v_0-\theta)e^{\kappa t}
	+\frac{1}{\kappa}\left(\frac{\theta}{2}-v_0\right).
\end{equation}

Assuming known values $\rho_{T_1}$ and $\rho_{T_2}$ for times $0<T_1<T_2$, we want to determine the forward correlation coefficient $\rho_{T_1,T_2}$ to be  active between times $T_1$ and $T_2$ such that the covariance between  the Brownian motions of the variance process and the exchange rate process agrees with the given values $\rho_{T_1}$ and $\rho_{T_2}$. This problem has a simple answer, namely: 
\begin{equation}
\rho_{T_1,T_2}=\rho_{T_2}, \ \ \ T_1\leq t\leq T_2.
\end{equation}
This can be seen by writing the Heston model in the form:
\begin{eqnarray}
dS_t&=&S_t \left( \mu\,dt+\sqrt{v_t}\,dW_t^{(1)} \right), \\
dv_t&=&\kappa(\theta-v_t)\,dt +\rho\sigma\sqrt{v_t}\,dW_t^{(1)}+\sqrt{1-\rho^2}\sigma\sqrt{v_t}\,dW_t^{(2)},
\end{eqnarray}
for a pair of independent Brownian motions $W^{(1)}$ and $W^{(2)}$. Observe that choosing the forward correlation coefficient as stated does not conflict with the computed forward volatility.

\xsubsection{Jump-diffusion models}{hes_JD}{}
\index{jump-diffusion model}

While trying to calibrate short term smiles, the volatility of volatility often seems to explode along with the speed of mean reversion. This is a strong indication that the process `wants' to jump, which of course it is not allowed to do. This observation, together with market crashes, has lead researchers to consider models with jumps \cite{gat:06,mar:sen:02}. Such models have been investigated already in the mid-seventies \cite{mer:76}, long before the advent of SV. Jump-diffusion (JD) models are, in general, more challenging to handle numerically than SV models. Like the latter, they result in an incomplete market. But, whereas SV models can be made complete by the introduction of one (or a few) traded options, a JD model typically requires the existence of a continuum of options for the market to be complete.

\citeasnoun{bat:96} and \citeasnoun{bak:cao:che:97} suggested using a combination of jumps and stochastic volatility.\index{jump-diffusion model!with stochastic volatility} This approach allows for even a better fit to market data, but at the cost of a larger number of parameters to calibrate from the same market volatility smile. \citeasnoun{and:and:00} let the stock dynamics be described by a JD process with local volatility.\index{local volatility} This method combines ease of modeling steep, short-term volatility skews (jumps) and accurate fitting to quoted option prices (deterministic volatility function). Other alternative approaches utilize L\'evy processes \cite{con:tan:03,ebe:kal:kri:03} or mixing unconditional disturbances \cite{tom:dec:06}.

\SKparam{SBIBLIOGRAPHY}{}

\end{document}